\documentclass[conference]{IEEEtran}
\IEEEoverridecommandlockouts
\usepackage[table]{xcolor}
\usepackage{cite}
\usepackage{amsmath,amssymb,amsfonts}
\usepackage{algorithmic}
\usepackage{graphicx}
\usepackage{textcomp}
\usepackage{xcolor}
\usepackage{tipa}
\usepackage{arabtex, lipsum}
\setfarsi

\usepackage{utf8} 
\setcode{utf8}

\usepackage{pifont}

\usepackage{ifxetex}
\usepackage{graphicx}

\usepackage[font=small,skip=0pt]{caption}
\usepackage{adjustbox}

\newcommand{\cmark}{\ding{51}}%
\newcommand{\xmark}{\ding{55}}%
\def\BibTeX{{\rm B\kern-.05em{\sc i\kern-.025em b}\kern-.08em
    T\kern-.1667em\lower.7ex\hbox{E}\kern-.125emX}}

\usepackage{etoolbox}
\makeatletter
\patchcmd{\@makecaption}
{\scshape}
{}
{}
{}
\makeatother

\newcommand \TableRef[1]{Tab. \ref{#1}}

\begin{document}

\title{ParsiNorm: A Persian Toolkit for Speech Processing Normalization\\
}

%

\author{%
	\IEEEauthorblockN{%
		Romina Oji\IEEEauthorrefmark{1},
		Seyedeh Fatemeh Razavi\IEEEauthorrefmark{1}\IEEEauthorrefmark{2}\IEEEauthorrefmark{3},
		Sajjad Abdi Dehsorkh\IEEEauthorrefmark{1}\IEEEauthorrefmark{2}\IEEEauthorrefmark{3},
	\\
		Alireza Hariri\IEEEauthorrefmark{1}\IEEEauthorrefmark{2},
		Hadi Asheri\IEEEauthorrefmark{1}\IEEEauthorrefmark{2} and
		Reshad Hosseini\IEEEauthorrefmark{1}\IEEEauthorrefmark{2}%
		\\
		\{romina.oji, razavi\_f, sadjad.abdi, alireza.hariri, hadi.asheri, reshad.hosseini\}@ut.ac.ir
	}%
	\IEEEauthorblockA{\IEEEauthorrefmark{1} School of ECE, College of Engineering, University of Tehran, Tehran, Iran}%
	\IEEEauthorblockA{\IEEEauthorrefmark{2} Hoosh Afzare Rahbare Aryaman (HARA) Company, Tehran, Iran}
	\IEEEauthorblockA{\IEEEauthorrefmark{3}Equal contribution}%
}

\maketitle

\begin{abstract}
In general, speech processing models consist of a language model along with an acoustic model.
Regardless of the language model's complexity and variants, three critical pre-processing steps are needed in language models: cleaning, normalization, and tokenization.
Among mentioned steps, the normalization step is so essential to format unification in pure textual applications.
However, for embedded language models in speech processing modules, normalization is not limited to format unification.
Moreover, it has to convert each readable symbol, number, etc., to how they are pronounced.
To the best of our knowledge, there is no Persian normalization toolkits for embedded language models in speech processing modules,
So in this paper, we propose an open-source normalization toolkit for text processing in speech applications.
Briefly, we consider different readable Persian text like symbols (common currencies, \#, @, URL, etc.), numbers (date, time, phone number, national code, etc.), and so on.
Comparison with other available Persian textual normalization tools indicates the superiority of the proposed method in speech processing.
Also, comparing the model's performance for one of the proposed functions (sentence separation) with other common natural language libraries such as HAZM and Parsivar indicates the proper performance of the proposed method.
Besides, its evaluation of some Persian Wikipedia data confirms the proper performance of the proposed method.
\end{abstract}


\section{Introduction}
{\color{black}
Pre-processing is one of the essential steps in the applications of Natural Language Processing (NLP) and Speech Processing (SP) \cite{jurafski}.
Normalization is the main component of pre-processing that leads to having a standard format for the text.
In this regard, some popular libraries like Spacy\footnote{https://spacy.io/} and Nltk\footnote{https://www.nltk.org/} have been presented.
However, the accuracy of these methods for Persian language is relatively low due to limited resources, while this problem is critical in the Persian language.
Because, some characteristic of the Persian language leads to appear some challenges in pre-processing.
Although some may not be the case for standard texts like news, it will be seen a lot for crawled texts containing conversational and colloquial sentences.
Since, the grammar of formal sentences is somewhat different.
For example, the following two sentences have the same meaning but do not have the same spelling rule:
"\<آيا به كتابخانه برويم؟>"
and
"\<بريم كتابخونه؟>".
Furthermore, some words are written in several ways like
"\<مساله>",
"\<مسأله>",
and
"\<مسئله>".
The presence of symbols from other languages such as Arabic and English, emojis, punctuations, etc., make more challenges for the normalization step.
In addition to the mentioned points, the normalization step will be more challenging in speech modules when we integrate a language model in SP.
Since a language model is suitable, that is close to the human reading of sentences.
For example, reading a phone number varies according to separating the number of digits in different people.
For example,
"1234"
is pronounced in many forms like
"\<دوازده سی و چهار>",
"\<هزار و دويست و سی وچهار>",
etc.
Therefore, besides mentioned normalization, some other operators need to be added for SP normalization purposes.

All in all, there are not many dedicated libraries for the Persian language and it suffers from a lack of coherent tools. The most famous of them are HAZM\footnote{https://www.sobhe.ir/hazm/} and Parsivar \cite{parsivar}.
On the other hand, the focus of the available tools has been solely on textual applications.
Therefore, there is no tool in this language that is efficient in the combined applications of text and audio.
In other words, the pre-processing method should cover the preprocessing steps of SP in addition to NLP.
Best of our knowledge, It has not been developed any tools with this consideration.
Therefore, we develop a dual-purpose tool that includes both NLP and SP normalization\footnote{Our open-source implementation is available at https://github.com/haraai}.
Our main contributions are as follows:
}
{\color{black}}
\begin{itemize}
	\item Improving existing general normalization (NLP normalization)
	\item Developing a new algorithm for sentence separation.
	\item Developing a new SP normalization presented for the first time in the Persian language
	\item Randomly generating different spelling for non-standard words like symbols, time, etc
\end{itemize}

{\color{black}
The paper is organized as follows:
Sec.\ref{related_works} provides a brief overview of previous methods.
Sec.\ref{proposed_method} describes details of the proposed method.
Experiments and comparisons with other available pre-processing toolkits are presented in Sec.\ref{experiment}
Finally, the paper is concluded in Sec.\ref{conclusion}.
}

\section{Related works}
\label{related_works}

{\color{black}
Text pre-processing is the primary step in NLP and SP applications.
Three essential components in this stage consist of cleaning, normalization, and tokenization.
Developing an SP normalization is the primary goal of this paper.
Therefore, there is more attention on normalization and related research.
In the Persian language, the main focus of researchers has been on text normalization.
Therefore, no suitable general tool for text normalization in embedded audio models has been provided, while there are different Persian textual tools.
Therefore, briefly, an overview of textual normalizations for the Persian language is presented.

One of the open-source libraries offered for NLP normalization is called HAZM.
Normalization and pos-tagging are some of its possibilities.
Empirical evidence from the use of this library suggests that despite the library's capability, the results are not necessarily accurate in some situations.
Among the available tools for the Persian language, only Parsivar seems to be a significant competitor for our work.
However, its focus has also been on pre-processing the text and unifying the text. 
The most important things covered in its normalization are converting numbers to their Persian text form, removing extra spaces, and removing punctuation marks related to other languages such as Arabic.
Although Parsivar also has instructions for normalization for dates, the important thing is that the way it is processed is not the same as how humans read it.
Stemming and lemmatization are other modules in parsivar that make it suitable for textual processing. However, they are not more needed in the SP normalization anymore.
Sentence segmentation is another ability of Parisivar that acts based on the end-of-sentence punctuations.
While it is not always true.
Because there are many exceptions in the Persian sentences like "..."  or existing abbreviations separated by ".".
Another critical point is the conversational sentences that are not necessarily distinguishable from the punctuation marks, and the grammar of those sentences is different from the formal sentences.
\\
Besides, the focus of all tools has been limited to textual applications.
Meanwhile, due to the importance of SP normalization \cite{jurafski}, we introduce ParsiNorm that performs normalization related to SP modules.
}
\section{Proposed method}
\label{proposed_method}
ParsiNorm is a tool written in python, and it is suitable for Persian text normalization for text and speech applications.  Before every SP system, a text normalization pre-processing is needed for handling non-standard words such as numbers, dates, time, etc \cite{jurafski}. Each non-standard word must be written how it is pronounced. For example, room number 101 is pronounced (one oh one), or the year 2021 is pronounced twenty twenty-one, so different numbers have different reading styles based on their types, both in Persian and English. We have developed two different normalizations. Sec. A explains one of them, which can be used for every Persian application using text. Sec. B explains the other normalization which is used as pre-processing for SP systems. We have investigated Persian language features and developed functions that converts words with different reading pronunciation and writing style to their reading forms.
\subsection{\textbf{General Normalization}}
Near all raw texts need some normalization. In this section, we explain different general implemented normalizations that can be used for Persian texts regardless of what applications these texts are going to be used.
\begin{itemize}	
	\item \textbf{Same character encoding:}
	The Persian language has the same character shapes as Arabic, but there are some differences. For example, the character 
	"\<ی>"
	 has different written forms in Arabic, like
	 "\<ے>",
	"\<ي>".
	we convert all of the Persian alphabets to the same format. It is common to write some English words in Persian writings like some special names, but the problem is that people do not use the same Character format, so we convert all English characters to the same form. For example, the letter "i" can be written in different formats like
	"\"i",
	"\textcircled{i}", etc. we convert all 26 English characters (Capital and small letters) to the same form.
	
	\item \textbf{Same Persian numbers:}
	There are two different formats for Persian numbers. We replace all of them with a unique form. Also, different English number formats are used in Persian texts, so we find and replace them with the same Persian forms.
	For example, number six in Persian has two different writing forms.
	Also, number six in English has various formats like "6", "\textbf{6}", 
	"\textcircled{6}",
	"\ding{177}",
	"\ding{187}",
	"\ding{197}",
	"\ding{207}",
	etc. we find them in sentences and replace them with Persian number "\<۶>".
	
		
	\item \textbf{Convert English and Arabic symbols to text:}
	Both in Persian and Arabic, some symbols are used instead of writing the whole word. For instance, some authors prefer to write symbol
	"\<صلّـے>"
	instead of
	"\<صلی>"
	, because it is more convenient. We find all of these symbols and convert them to whole words.
	
	\item \textbf{Normalize punctuations:}
	Some punctuations have different formats. We find all forms of them and convert them to the same format. For example, "\%" can be written in different Unicode such as
	"$^c/_o$" and "$^./_.$".
	
	\item \textbf{Convert Math symbols to Persian numbers:}
	Some math symbols such as
	"{\textonehalf}"
	are used in the Persian corpus. But the correct format is
	"\<۱>\</>\<۲>"
	which is a complete writing form with Persian numbers.
	
	\item \textbf{Replace HTML tags with text:}
	Because some Persian data sources are crowded web texts, there are some HTML tags in these texts.
	These tags are replaced with the correct and readable form.
	For example, less than HTML tag "\&lt" is replaced by "$<$". 
	
	\item \textbf{Remove emojis:}
	In Some tasks, emojis are not important, and it is better to be cleaned. We have gathered a large list of emojis and remove them from sentences.
	
	\item \textbf{Separate sentences:}
	Separating sentences at correct positions is an important task that can improve some text models. For instance, one of the tasks in the Bert language model is next sentence prediction, so it is important to separate each sentence at its correct position. In Persian, all sentences end with a dot, like English sentences. We can split sentences with dot marks, but there is a big problem: dots can be used in other parts of the sentence, such as abbreviations, floating-point numbers, emails, URLs, etc. We have gathered an extensive list of abbreviations from different categories and replace them with how they are read. Also, emails and URLs are converted to how they are pronounced. More information about abbreviations, emails are described in part SP Normalization. Floating-point numbers with dot splitters have a different mechanism that separates them from other dots in the sentence.
	
	Parsivar just considers dot floating points, but for the rest words with punctuation, it split sentences wrongly. We have improved Parsivar sentence separation and add the aforementioned features. This sentence separation considers that all sentences have a dot at the end, But some people forget to put a dot at the end of each sentence.	
	We use other Persian language features to separate sentences. In Persian formal sentences, verbs come at the end of the sentence. Using this feature, we can separate sentences by Verbs. To do this, we use the HAZM\footnote{https://www.sobhe.ir/hazm/} POS tagger to find the verb in a sentence, and then we separate each paragraph into some sentences.
\\
		
\end{itemize}

\subsection{\textbf{Speech Processing Normalization}}

Following normalization is used for SP applications. Some symbols, numbers, and texts are read differently from how they are written, so the spelled form must be replaced in speech applications.
\begin{itemize}
	
	\item \textbf{Replace all symbols with text:}
			Some symbols are used in both Persian and English texts. We have gathered a list of these symbols and replace them with how it is read. Some of these symbols are: star(*), square(\#), degree($^\circ$), percent(\%), and etc.
				 
	\item \textbf{Replace common currency with their readable text format:}
			Currency symbols are used instead of words. When everyone reads these symbols, tell the correct word of it. For example, the symbol \$ is pronounced dollar in English. We have gathered currency signs used in different countries and replaced their symbols to how they are read.

	\item \textbf{Replace Math symbols with Persian readable text:}
			As mentioned in the general normalization section, some math symbols are used in the Persian corpus. In general normalization, we replace them with Persian Numbers but, they have different reading formats. All of these symbols are converted to how they are read.
			
	\item \textbf{Replace abbreviations with how they are pronounced:}
			Both English and Persian abbreviations are used in the Persian corpus. Persian Abbrevians are read differently from how they are written, but English abbreviations are somehow read in the same format as they are written.
			When English abbreviations are read, we pronounce each letter separately. For example, the abbreviation Ph.D. is read
			"pi e\textsci t\textesh\  di"
	("\<پی‌اچ‌دی>").	
			We Find all English abbreviations (can end with a dot or not) and replace each English letter with how they are pronounced in Persian.
			Persian native speakers read abbreviations how they are written entirely Because all Persian abbreviations stand for the first letter of a long-phrase that contains multiple words. For example,
		"\<ر.ک>"
			 is the abbreviation of the phrase
			 "\<رجوع كنيد>"
			  which is used the first letter of this phrase
			  "\<ر>"
			   and
			   			   "\<ک>".
			    We have gathered an extensive list that contains about 200 abbreviations and their complete format. They are categorized into different categories such as law, books, date, time, and others. All abbreviations are replaced with their long-phrase form.
		
		\item \textbf{Replace URL and Emails to the text how it is pronounced:}
				Emails and URLs are pronounced differently from how they are written because some symbols such as "@", "-", "\_", etc. are a part of them. We replace these symbols with the words how they are pronounced. For instance, "@" is replaced by "at". Furthermore, we separate each part of emails and URLs and put a space between them because this is how they are read. Some URLs are very big when non-English characters are used and converted to other characters such as "\%", digits, and English alphabet. For these long URLs, we keep the main part of it and remove the sub-directory.
				For example "http://wpc.be1e.edgecastcdn.net/news/20ak9qy4prra.html" is converted to "http do noghte slash slash wpc dot be1e dot edgecastcdn dot net" by removing its sub-directories.
		
		\item \textbf{Convert date to text:}
			Dates can be written in two formats: text format like tenth January, twenty eighteen or numbers of the day, month, year, which are split by punctuations like "." , "-" and "/" for example, 10/1/2018. We find second forms and replace them with the first mentioned form. In Persian Corpus, three types of dates are used which are solar hijri, gregorian calendar, lunar hijri. They can be detected based on their years. We use some roles to detect the date type and replace them with how they are read. Detecting the type of date is important because the month names of these types are different from each other. For instance, the first month of solar hijri, gregorian calendar, lunar hijri 
			are in order: Farvardin, January, Moharam. We have made ten different templates that date can be read, and for each date, we select one of them randomly and replace the blanks with the appropriate form of day, month, and year. One example of converting solar hijri calendar date to text is converting
			1397/7/9
			to
			"\<نهم مهر سال هزار و سيصد و نود و هفت>".
		
		\item \textbf{Convert time to text:}
		Times are written in two specific forms. One of them is "hour:minute:second", and the other is "hour:minute". We replace both forms with the text format of how they are written. For Detection of time, we use some rules because some digits that are separated by ":" are not time. Different times are read differently. For instance, in English, the time 8:00 is read eight o\textquotesingle clock, and we do not read it eight zero while the time 8:25 is read eight twenty-five. We have investigated all ways that a time can be read and make several templates based on them. Then each blank of the templates is replaced with the appropriate part of the time.
		
	\item \textbf{Detect telephone Numbers:}
	Telephone numbers are read differently from other numbers. In Persian, people separate the telephone number into several parts that are three or two digits and then read each part. But some telephone numbers such as mobile phones have a beginning part which is always read in the same way. For example, the telephone number 0912332xxxxx  is a mobile number, and all people read the first four digits in the same way, which is "zero nine hundred twelve". We use several templates based on the type of phone number and its length and convert the telephone numbers to the text how it is read. In order to detect telephone numbers, we use several roles that a telephone number can be used in a text, for instance, the format of the number itself and the word before or after it. Then we use modified num2faword\footnote{https://pypi.org/project/num2fawords/} to convert separated numbers to text.
	
	\item \textbf{Convert special numbers to text:}
	Some numbers such as National Code, card number, Sheba are read differently from other numbers. Readers separate them to various parts that contain two or three digits and then read each part. We use Persian-Tools\footnote{https://pythonrepo.com/repo/persian-tools-py-persian-tool} to detect these numbers and num2faword to convert each piece of separated numbers to text. Also, generally, people do not read long numbers as to how they are. They do the same strategy for these numbers. So we Convert numbers with more than 15 digits to separated parts and convert each element to text.
		
\end{itemize}


%
%
%
%
%



\section{Experiment}
\label{experiment}

In this section, studies about the performance of the developed toolbox are presented.
At first, a comparison is made on the possibilities of the proposed method for improving NLP normalization along with proposing a new one for SP with other available text-specific normalization toolboxes in Persian.
Then, the performance of the proposed model is presented in the face of numerical samples (such as date, time and etc.).
After that, numerical analysis between the three methods (ParsiNorm, HAZM, Parsivar) is evaluated for a similar function (sentence separation).
Finally, the performance of the proposed model is shown on some extracted sentences from Wikipedia.

Generally, non-standard words such as numbers, monetary amounts, dates, and other concepts spelled differently than verbalized, need special pre-processing in SP systems \cite{jurafski}.
Based on Persian language features, we have converted various non-standard words to how they are pronounced.
As far as we are concerned, no publicly available tool for Persian SP normalization is available. We have compared ParsiNorm implemented functions with other normalization tools such as Hazm, Parsivar, persian-tool. The result of this comparison is available in \TableRef{tab1}.

\begin{table}[]
	\color{black}
	\caption{Comparison among available text-specific Persian toolboxes with ParsiNorm.}
	\begin{center}
		\resizebox{9cm}{!}
		{
			\begin{tabular}{|l|c|c|c|c|}
				\hline
				
				\rowcolor{lightgray}
				\textbf{}&\multicolumn{4}{|c|}{\textbf{Toolkits}} \\
				\cline{2-5}
				\rowcolor{lightgray}
				\textbf{Semiotic class} & \textbf{\textit{ours}}& \textbf{\textit{Parsivar}}& \textbf{\textit{HAZM}} & \textbf{\textit{persian-tools}} \\
				\hline
				English Abrreviation & \cmark & \xmark & \xmark & \xmark  \\
				Abbreviation of different topics & \cmark & \xmark & \xmark & \xmark \\
				\hline
				acronyms read as letters & \cmark & \xmark & \xmark & \xmark  \\
				\hline
				times & \cmark & \xmark & \xmark & \cmark  \\
				dates & \cmark & \cmark & \xmark & \xmark  \\
				\hline
				Same character encoding & \cmark & \cmark & \cmark$^1$ & \xmark  \\
				Same Persian numbers & \cmark & \cmark$^2$ & \cmark$^2$ & \xmark  \\
				\hline
				(English/Arabic) symbols to text & \cmark & \xmark & \xmark & \xmark  \\
				Math symbols to Persian numbers & \cmark & \xmark & \xmark & \xmark  \\
				Replace all symbols with text & \cmark & \xmark & \xmark & \xmark \\
				Replace currencies with text & \cmark & \xmark & \xmark & \xmark  \\
				\hline
				Numbers to text & \cmark & \cmark & \xmark & \cmark  \\
				Detect telephone numbers & \cmark & \xmark & \xmark & \cmark$^3$ \\
				Convert national-code to text & \cmark & \xmark & \xmark & \cmark$^3$  \\
				Convert card-number to text & \cmark & \xmark & \xmark & \cmark$^3$  \\
				Convert Sheba to text & \cmark & \xmark & \xmark & \cmark$^3$  \\
				\hline
				Replace HTML tags with text & \cmark & \xmark & \xmark & \xmark  \\
				URL and Emails to text 
				& \cmark & \xmark & \xmark & \xmark  \\
				\hline
				Normalize punctuations & \cmark & \cmark & \xmark & \xmark  \\
				Remove emojis & \cmark & \xmark & \xmark & \xmark  \\
				\hline
				Separate sentences & \cmark & \cmark$^4$ & \cmark$^4$ & \xmark  \\
				\hline
				Pinglish to Persians & \xmark & \cmark & \xmark & \xmark  \\
				\hline
				Informal to formal & \xmark & \xmark & \cmark & \xmark  \\
				\hline
				Replace spaces by half-space & \xmark & \cmark & \cmark & \cmark  \\
				\hline
			\end{tabular}
		}
		\label{tab1}
	\end{center}
	\footnotesize{$^1$Hazm converts Only two letters,
		"\<ک>" 
		and
		"\<ی>"
		from Arabic to Persian form.}\\
	\footnotesize{$^2$Only Arabic numbers are converted to Persian numbers, and the various form of numbers which can be the symbol of numbers such as \ding{174} are not converted to Persian numbers.}\\
	\footnotesize{$^3$The persian-tools just detect the type of numbers such as Telephone,Sheba, card number and national code. These numbers do not turn into the equivalent text of its readings.}\\
	\footnotesize{$^4$Sentence separating in Hazm and Parsivar is based on punctuations. This is not a good way of separating sentences because these punctuations can be used in some words that are not at the end of the sentence. So this way of separating sentences has lots of problems. Parsivar detects floating points and does not consider these dots as dots that come at the end of the sentences, But other types of dots do not indicate the end of sentences, such as dots used in abbreviations.}\\
\end{table}

The results show the superiority of ParsiNorm compared to other available methods for SP applications.



{\color{black}

On the other hand, different numbers have different reading formats. For example, long numbers and special numbers (such as phone numbers, national codes, banking card numbers, etc.) are read differently from written ones. Also, times and dates are read differently. For instance, the time "10:30:25", we do not express ":" in daily speaking but define changes from hour to minutes by using special characters such as
"\<و>"
or telling that what minute and second are
"\<ده و سی دقيقه و بيست و پنج ثانيه>".
Therefore, we consider all ways a number (like date, time, etc.) can be read.
Finally, one of them is selected randomly and is replaced with the written format. We have shown different ways of telling each number in \TableRef{tab2}.
As seen, we do not necessarily have a specific output, and each particular number can have a different result every time.

\begin{table}[]
	\color{black}
	\caption{Some examples of the performance of the proposed method in the face of numerical samples with considering different pronunciations.}
	\begin{center}
		\resizebox{9cm}{!}
		{
		\begin{tabular}{c}
			\hline
			\textbf{examples} \\
			\hline
			\cellcolor{gray!25}
			Semiotic class: \textbf{times} \\ 
			\cellcolor{gray!15}
			input: 11:35 \\
			\cellcolor{gray!5}
			\<يازده و سی و پنج>
			\\
			\cellcolor{gray!5}
			\<يازده و سی و پنج دقيقه>
			\\
			\cellcolor{gray!25}
			Semiotic class: \textbf{dates} \\ 
			\cellcolor{gray!15}
			input: 1400-07-25 \\
			\cellcolor{gray!5}
			\<بيست و پنج  مهر ماه هزار و چهارصد>
			\\
			\cellcolor{gray!5}
			\<بيست و پنجم مهر هزار و چهارصد>
			\\
			\cellcolor{gray!5}
			\<بيست و پنج مهر سال هزار و چهارصد>
			\\
			\cellcolor{gray!5}
			\<بيست و پنج هفت هزار و چهارصد>
			\\
			\cellcolor{gray!25}
			Semiotic class: \textbf{detect telephone numbers and convert} \\
			\cellcolor{gray!15}
			input: 09397796915 \\
			\cellcolor{gray!5}
			\<صفر نهصد و سی و نه هفتاد و هفت نود و شش نهصد و پانزده>
			\\
			\cellcolor{gray!5}
			\<صفر نهصد و سی و نه هفتاد و هفت نهصد و شصت و نه پانزده>
			\\
			\cellcolor{gray!5}
			\<صفر نهصد و سی و نه هفتصد و هفتاد و نه شصت و نه پانزده>			
			\\
			\cellcolor{gray!25}
			Semiotic class: \textbf{detect persian national code and convert to text} \\ 
			\cellcolor{gray!15}
			input: 0523924984 \\
			\cellcolor{gray!5}
			\<صفر  پنج   بيست و سه   نود و دو   چهل و نه   هشتاد و چهار>
			\\
			\cellcolor{gray!5}
			\<صفر  پنجاه و دو   سی و نه   دويست و چهل و نه   هشتاد و چهار>
			\\
			\cellcolor{gray!25}
			Semiotic class: \textbf{detect card number and convert to text} \\
			\cellcolor{gray!15}
			input: 6104337852441441 \\
			\cellcolor{gray!5}
			\<شصت و يک صفر  چهار   سی و سه   هفتاد و هشت   پنجاه و دو>
			\\
			\cellcolor{gray!5}
			\<چهل و چهار   چهارده   چهل و يک>
			\\
			\hline			 
		\end{tabular}
	}
		\label{tab2}
	\end{center}
\end{table}

Furthermore, the performance of one of the functions is compared with other methods.
Breaking long sentences into shorter ones is usually done by identifying the '.' in the sentence by HAZM or Parsivar.
HAZM, however, also takes into account several modes of abbreviation.
But the proposed method is not limited to the point but also considers the maximum possible scenarios. Because in Persian, '.' does not necessarily mean the end of the sentence and is used in other cases such as decimal, abbreviation, '.' in the URLs, etc.
Accuracy for the three methods on 107 random sentences from Wikipedia is presented in \TableRef{tab3}.

\begin{table}[]
	\color{black}
	\caption{The accuracy of ParsiNorm, HAZM, and Parsivar for 107 random sentences from Wikipedia for sentence separation.}
	\begin{center}
		\resizebox{9cm}{!}
		{
			\begin{tabular}{|l|c|c|c|}
				\hline
				\rowcolor{lightgray}
				\textbf{}&\multicolumn{3}{|c|}{\tiny \textbf{Toolkits}} \\
				\rowcolor{lightgray}
				\textbf{\tiny Criterion} & \textbf{\tiny \textit{ParsiNorm}}& \textbf{\tiny \textit{HAZM}}& \textbf{\tiny \textit{Parsivar}} \\
				\hline
				\tiny Accuracy & \textbf{\tiny 89.71} & \tiny 88.78 & \tiny 69.15  \\
				\hline
			\end{tabular}
		}
		\label{tab3}
	\end{center}
\end{table}

To further intuition, the performance of the proposed method on formal some extracted sentences from Persian Wikipedia is depicted in Fig. 1. 

\begin{figure}[]
	\label{fig}
	\begin{adjustbox}{addcode={\begin{minipage}{\width}}{\caption{%
						The performance of ParsiNorm on some sentences include numbers and abbreviations from Persian Wikipedia.
			}\end{minipage}},rotate=90,center}
		\includegraphics[scale=.6]{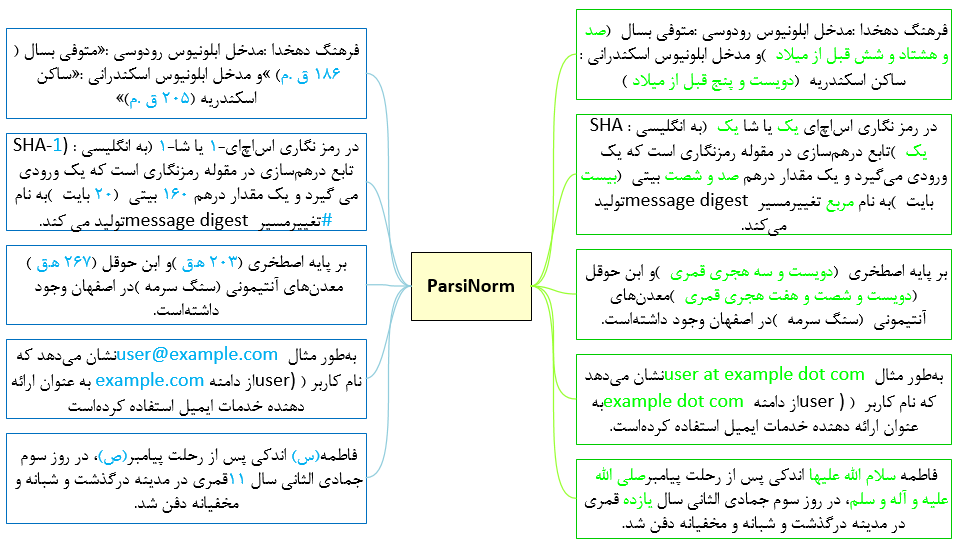}%
	\end{adjustbox}
\end{figure}

}

\section{Conclusion}
\label{conclusion}

This paper proposed ParsiNorm toolkit, an open-source normalization tool for the Persian language in both text and speech processing applications. 
Randomly generating a different reading form for non-standard words like time, symbol, number, etc., is the main contribution in this paper.
Furthermore, proposing a robust Persian sentence recognition method is another one.
Comparison with available basic libraries, along with the presented results about the method's performance on some of the sentences, indicates the excellent performance of the proposed method.

%

%



\end{document}